\documentclass[preprint,12pt]{elsarticle}

\usepackage{amssymb}
\usepackage{graphicx}
\usepackage{color,soul}
\usepackage{subfigure}

\begin{document}
\begin{frontmatter}


\title{
\textbf{Unraveling the effects of anionic vacancies and temperature on mechanical properties of NbC and NbN: Insights from Quantum  Mechanical  Study} }


\author[inst1]{P.W. Muchiri}

\affiliation[inst1]{organization={Materials Modeling Group, Department of Physics, Earth and Environmental Sciences, The Technical University of Kenya},
            addressline={P.O. Box, 52428}, 
            city={Nairobi},
            postcode={00200}, 
            country={KENYA}}

\author[inst2]{K. K. Korir}
\affiliation[inst2]{organization={Department of Mathematics, Physics, and Computing, Moi University},
            addressline={P.O BOX 3900}, 
            city={Eldoret},
            postcode={30100}, 
            country={Kenya}}
\author[inst3]{N. W. Makau}
\affiliation[inst3]{organization={Department of Physics, University of Eldoret},
            addressline={P.O BOX 1125}, 
            city={Eldoret},
            postcode={30100}, 
            country={Kenya}}        
\author[inst1]{M. O. Atambo}
\author[inst1]{G. O. Amolo}

\begin{abstract}

Transition metal carbides and nitrides (TMCNs), such as niobium carbide (NbC) and niobium nitride (NbN), are of great technological interest due to their exceptional hardness, high melting points, and thermal stability. While previous studies have focused on their ground-state properties (at 0 K), limited information exists on their mechanical behavior under realistic operational conditions involving elevated temperatures and the presence of defects. In this study, we employ \textit{ab initio} molecular dynamics (AIMD) simulations to investigate the effects of temperature (300–1500 K) and anionic vacancies (NbC/N$_{0.7500–0.9375}$) on the mechanical properties of NbC and NbN in rocksalt (RS), zincblende (ZB), and wurtzite (WZ) structures. The results reveal a nonlinear decrease in elastic constants, bulk, shear, and Young’s moduli with both increasing temperature and defect concentration. Hardness and toughness analyses, based on Pugh’s ratio and Poisson’s ratio, show ductility–brittleness transitions that are sensitive to structure, defect level, and thermal effects. Furthermore, vacancy migration energies computed using the nudged elastic band (NEB) method demonstrate strong structural dependence, with RS exhibiting the highest barriers and WZ the lowest. These findings provide new insights into the defect–temperature interplay in NbC and NbN, offering guidelines for their optimization in high-temperature and wear-resistant applications.

\end{abstract}




\end{frontmatter}


\section{Introduction}
The demand for advanced materials with tailored properties for energy, mining, construction, and electronics has necessitated the development of engineered materials. Indeed, naturally occurring materials often suffer from limited performance, high costs, property variability, and restricted availability. To address these challenges, researchers use a variety of design and optimization tools, including computer simulations, to develop next-generation designer materials.
Refractory materials such as Group IV-VI transition metal carbides and nitrides (TMCNs) are vital for ultra-hard industries applications i.e machining, which is attributed to novel properties. These exceptional properties include high melting points and hardness, wear resistant, chemical inert and high thermal and electrical conductivity \cite{oyama1996introduction, tothtransition}. For example, early transition metal nitrides such as TiN, CrN, HfN and ZrN were found to be ideal for industrial cutting tools and wear resistant coatings \cite{dongil2019recent}. 
Other  studies have demonstrated that TMCNs possesses high stiffness constants and resistance capabilities thus rendering them ideal for high temperature applications in fields such as aerospace, power generation and automobile industry \cite{clemens2011light, clemens2013design,kim2014effects, feng2014influence}.  Ni-based alloys have been shown to perform reliably at elevated temperatures around 730–900°C, with modern alloys like René 77 exhibiting excellent micro-structural stability and high tensile strength \cite{grudzien2025solidification}. Recent advances in alloy design, heat treatment, and manufacturing have significantly improved the performance of these materials, allowing their use in critical applications under extreme thermal and mechanical conditions.

 Synthesis, post-growth treatment, and device operation occurs at elevated temperature (up to 1500 °C  in case of Ni-based alloys), therefore, structural and mechanical properties are projected to be significantly influenced. However, the effects of temperature on the crystal micro-structure and associated properties are seldomly studied, yet their effects have profound impact on the performance of such device. For instance, some of the common mechanical properties such as hardness and toughness are calculated based on elastic constants, shear and bulk modulus obtained at ground state (0 K) \cite{korir2011first, muchiri2019hardness} and likely to give inaccurate picture of elastic regime. The state of the art density functional theory (DFT) calculations simulate static models with atom fixed at specific lattice points and at 0 K. This approach usually neglect zero-point atomic motion and as a result there are non-negligible errors in some parameters \cite{isaev2011impact}.

Further, the behavior of defects such as vacancies at high temperature and their effect on mechanical properties is critical for comprehensive evaluation of device performance. For example, defects are known to degrade the mechanical robustness of TMCNs at 0 K \cite{muchiri2022impact, jhi1999electronic} but   rigorous analysis at elevated temperature is lacking. It is widely acknowledged that materials even in their pristine state are not devoid of defects \cite{koutna2021high, de2015charting}. Thus, parameters that are essential in evaluating mechanical response of materials such as shear modulus, Poisson ratio, hardness and toughness should be determined with respect to temperature \cite{balasubramanian2018valence,edstrom2014effects,abadias2014alloying,mazhnik2020application}. 
During utilization of TMCNs as machining tools and related applications, the temperature may rise above 1000 K, thus such devices are envisaged to operate at high temperature. This makes it crucial to put into account the finite temperature effects on the theoretical values of elastic constants, bulk, shear and Young’s modulus.
The initiative to design ceramic coatings that possess improved hardness has been the recent focus of researchers \cite{veprek2005different,shin2003vacancy,jhi1999electronic}.  Indeed, the operation of such devices entails turning and twisting forces which may not be ideal for brittle system thus the need to develop a robust system with extreme hardness and high tolerance to turning forces. Other desirable device traits include limited defect propagation, and enhanced thermal conduction. Therefore, this study explore the effect of temperature (0 – 1500 K) and defects (anionic vacancies) on the mechanical response of NbC and NbN.
Within the experimental set-up, it is often difficult to explore defect at atomic-scale and its implication on mechanical properties, instead, researchers relay on computer simulation such as ab initio Molecular Dynamics. For example, Yu et al \cite{yu2014vacancy} showed that increase in vacancy concentration in MgSi results in nonlinear decrease in stress. Similarly, Yang et al \cite{yang2012molecular} found that the presence of vacancy defects in CoSb$_{3}$  results in decrease in stress and elastic modulus with increase in temperature and vacancy concentration. However, other studies \cite{mayrhofer2003self} attributed age hardening to spinodal decomposition in transition-metal nitride alloys. While NbC and NbN have been identified as potential candidates for high-temperature applications, there are limited studies that explore the effects of vacancies and temperature on their mechanical response. The results of such investigations may pave the way for the development and optimization of devices derived from NbC and NbN, thus the thrust of the study.

\section{COMPUTATIONAL DETAILS}
The Vienna Ab Initio Simulation Package (VASP) \cite{kresse1993ab} was used to perform molecular dynamics. The exchange-correlation functional is approximated using generalized gradient approximation (GGA) as implemented in projected augmented wave method as per Perdew-Burke-Ernzerhof (PAW-PBE) \cite{blochl1994projector} with electronic convergence criterion of 10$^{-8} eV$. The calculations were performed using a supercell approach with cell sizes of \({2\times 2 \times 2}\) and \({2\times 2 \times 4}\), for NbC and NbN, respectively at Gamma point \cite{hull2011introduction}. 
Ab Initio molecular dynamics were performed for NbC and NbN in RS, ZB and WZ  structures for a temperature range between 300 K to 1500 K that correspond to operation temperature for the various applications of TMCNs. The various structures were initially equilibrated for 2 $fs$ at 300 K using NVT,  afterward the temperature was increased gradually to 1500 K using Nose-Hoover thermostat \cite{nose1984unified}  with a time-step of 2 $fs$. The overall MD simulation time step was between $5$ and $10$ ps with the atomic trajectory determined by numerical integration of equation of motion using Verlet velocity algorithm \cite{andersen1980molecular}.
Climb-image nudged elastic band \cite{jonsson1998nudged} was used to determine the transition state using tangent estimates and implemented the VTST version of VASP \cite{voter1985dynamical} and minimum of 5 images used to map the vacancy diffusion path.  

\section{RESULTS AND DISCUSSIONS}

Mechanical characterization parameters such as elastic moduli are essential for design and discovery of novel materials. For example, Bon Mises criterion is used to determine the mechanical stability of materials. This work explore stress-strain relations/dynamic in the presence of anionic vacancies. Indeed, Muchiri et. al. \cite{muchiri2022impact}  showed that defect/vacancies depletes the mechanical properties such as bulk, shear and Young’s modulus, elastic constants and Vicker’s hardness. Since such studies were  performed at ground state, the interplay of anionic vacancies and temperature is not accounted. At elevated temperatures, lattice deformation increases dramatically due to enhanced atomic motion and results in plastic deformation.

\subsection{\textbf{The Elastic Properties of NbC and NbN}}
The elastic constants obtained at room temperature (300 K) using MD approach agrees well with corresponding results obtained at ground state (0 K). Given the limited experimental studies on the temperature dependence of elastic constants of NbC and NbN, the reliability of the results obtained in this study is demonstrated through comparison with available experimental data \cite{jun1971elastic}. Shaffer et. al. \cite{jun1971elastic} determined the Young’s modulus of NbC for varying defect concentration, with NbC$_{0.95}$, NbC$_{0.964}$ and NbC$_{0.969}$ having 478 GPa, 488 GPa and 451 GPa, respectively. While shear modulus of  NbC$_{0.95}$, NbC$_{0.964}$ and NbC$_{0.969}$ were found to be 197 GPa, 199 GPa and 189 GPa, respectively. In this study, the mechanical properties were observed to decrease in magnitude with increasing temperature.  Here we report mechanical properties of NbC and NbN in RS, ZB and WZ  for a temperature range of 0 – 1500 K.

\begin{figure}[ht!]
\centering%
\includegraphics[width=1.1\linewidth]{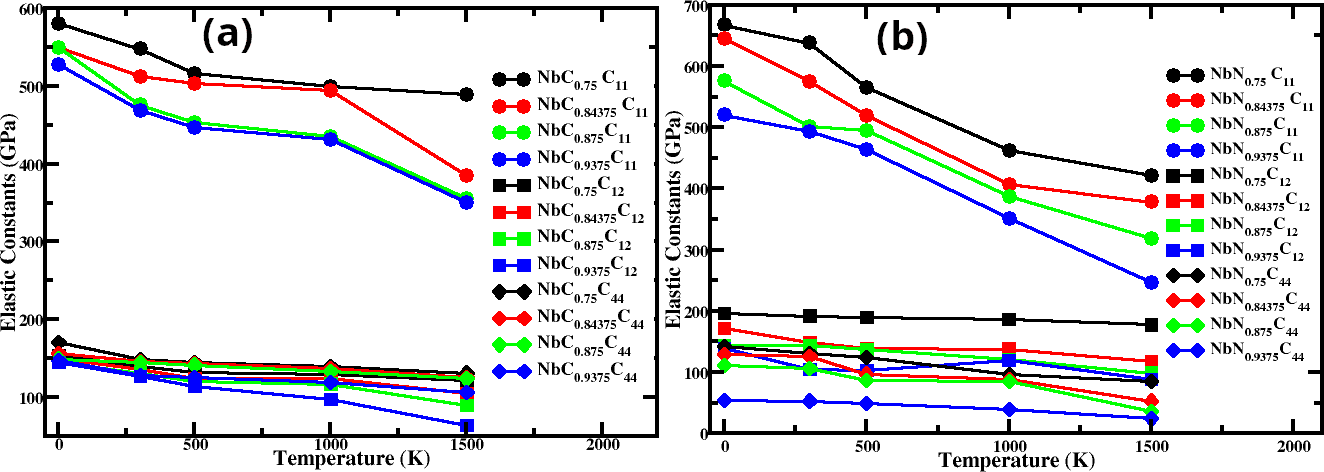}%
\caption{ Calculated elastic constants of (a) NbC and (b) NbN in RS crystal structure at different temperatures and defect concentrations. }
\label{NbC_N_C_RS.png}
\end{figure} 

\begin{figure}[ht!]
\centering%
\includegraphics[width=1.1\linewidth]{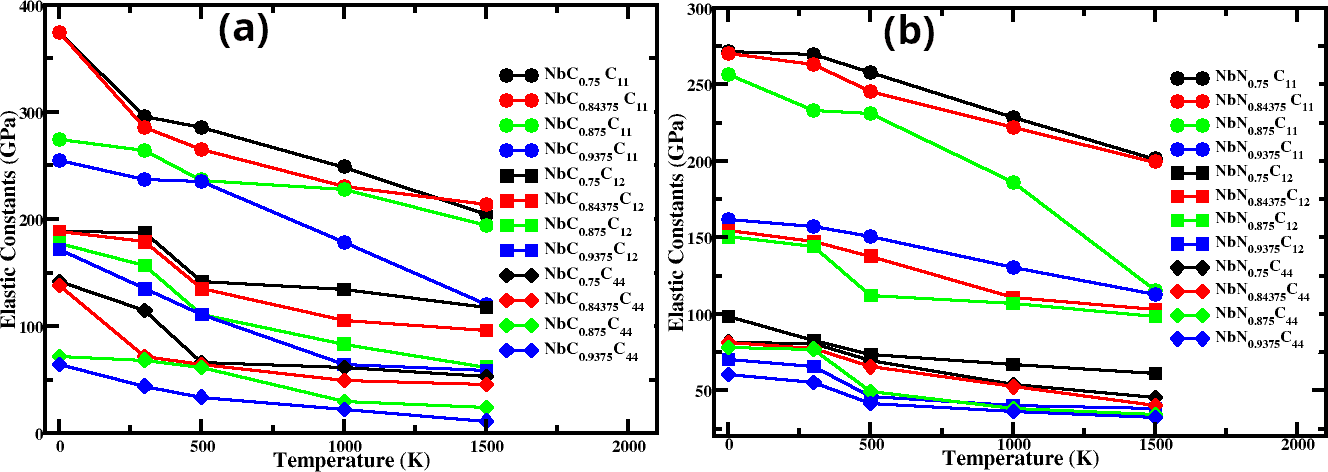}%
\caption{ Calculated elastic constants of (a) NbC and (b) NbN in ZB crystal structure at different temperatures and defect concentrations. }
\label{NbC_N_C_ZB.png}
\end{figure} 

\begin{figure}[ht!]
\centering%
\includegraphics[width=1.1\linewidth]{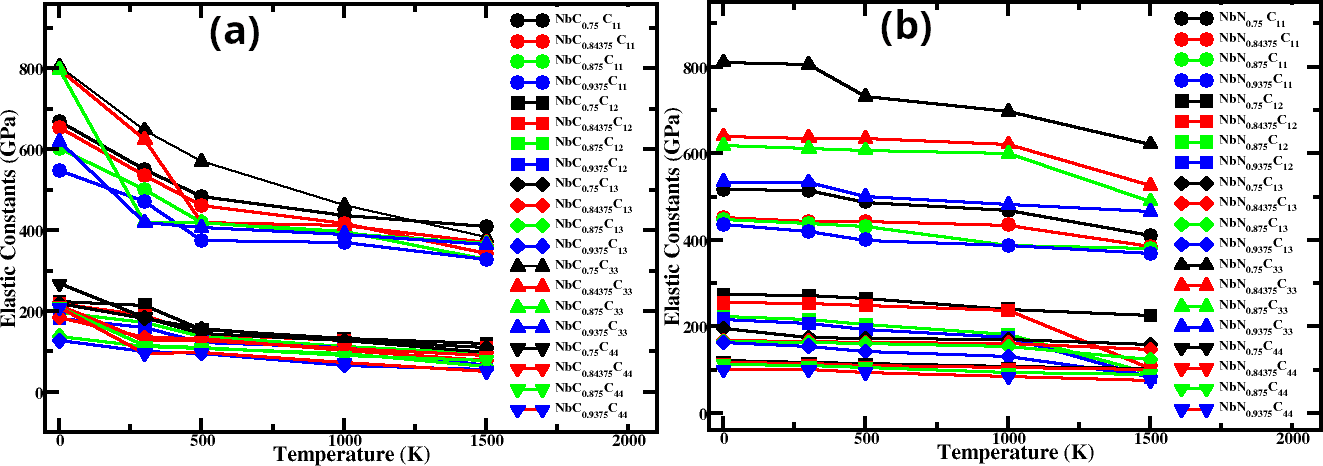}%
\caption{ Calculated elastic constants of (a) NbC and (b) NbN in WZ crystal structure at different temperatures and defect concentrations. }
\label{NbC_N_C_WZ.png}
\end{figure} 

In Figure \ref{NbC_N_C_RS.png}, the elastic constants C$_{11}$ is observed to reduce dramatically with increase in temperature. For example, temperature increase from 0 K to 1500 K yields a reduction of C$_{11}$ of up to 15.7 \% and 36.7 \% for NbC$_{0.75}$ and NbN$_{0.75}$, respectively, as shown in Figure \ref{NbC_N_C_RS.png}(a-b). Similarly, C$_{12}$ and C$_{44}$ tend to be  significantly affected by temperature and defect increase. For example, an increase in  temperature from 0 K to 1500 K leads to a decrease of C$_{12}$ by up to 21.0 \% and 9.3 \% for NbC$_{0.75}$ and NbN$_{0.75}$, respectively, as shown in Figure \ref{NbC_N_C_RS.png}.

In Figure \ref{NbC_N_C_ZB.png}, the elastic constant C$_{11}$ in ZB is also observed to reduce with increase in temperature. For instance, as the temperature rises from 0 K to 1500 K, this elastic constant tend to decrease by up to 45.4 \% and 25.8 \% for  NbC$_{0.75}$ and NbN$_{0.75}$, respectively, as shown in Figure \ref{NbC_N_C_ZB.png}(a-b). In addition, a change in  the values of C$_{12}$ and C$_{44}$ is noticeable with increase in temperature, where a 37.6 \% and 30.4 \% decrease is observed in the values of C$_{12}$ of NbC$_{0.75}$ and NbN$_{0.75}$, respectively, as shown in Figure \ref{NbC_N_C_ZB.png}(a-b).
Similarly, Figure \ref{NbC_N_C_WZ.png}(a-b) represents the effect of both defects and temperature on the elastic constants of NbC and NbN in WZ crystal structure. The elastic constants C$_{11}$, C$_{12}$, C$_{13}$, C$_{33}$ and C$_{44}$ decrease by up to 39.0 \%, 45.9 \%, 49.7 \%, 52.6 \% and 63.4 \%, respectively, for NbC$_{0.75}$ and 20.4 \%, 48.4 \%, 18.9 \%, 23.4 \% and 14.9 \%, respectively, for NbN$_{0.75}$. 
This trend  is also observed as the defect concentration increase with temperature rise. The values of elastic constants decreases with increase in defect and temperature as shown in Figures \ref{NbC_N_C_RS.png}-\ref{NbC_N_C_WZ.png}. The reduction in elastic constants can be attributed to decreased bonding due to increased thermal movements.

\subsection{\textbf{Mechanical properties of NbC and NbN}}
Mechanical properties determine material’s behaviour under external forces and inform their suitability for the specific applications such as the strength for load-bearing or toughness for impact resistance. The properties investigated in this study as a function of temperature include bulk modulus (B), Young’s modulus (E) and shear modulus (G). Bulk and shear moduli were determined using the Voight-Reuss-Hill method as expressed below,
 \begin{equation}\label{eq:bulk_modulus}
B = \frac{C_{11}+2C_{12}}{3}
\end{equation}

 \begin{equation}\label{eq:shear_modulus}
G = \frac{1}{2}(G_{V}+G_{R})
\end{equation} 

with  
 \begin{equation}\label{eq:GV}
G_{V} = \frac{C_{11}-C_{12} +3C_{44}}{4}
\end{equation}

\begin{equation}\label{eq:GR}
G_{R} = \frac{(C_{11}-C_{12})5C_{44}}{4C_{44}+3C_{11}-3C_{12}}
\end{equation}

\begin{figure}[ht!]
\centering%
\includegraphics[width=1.1\linewidth]{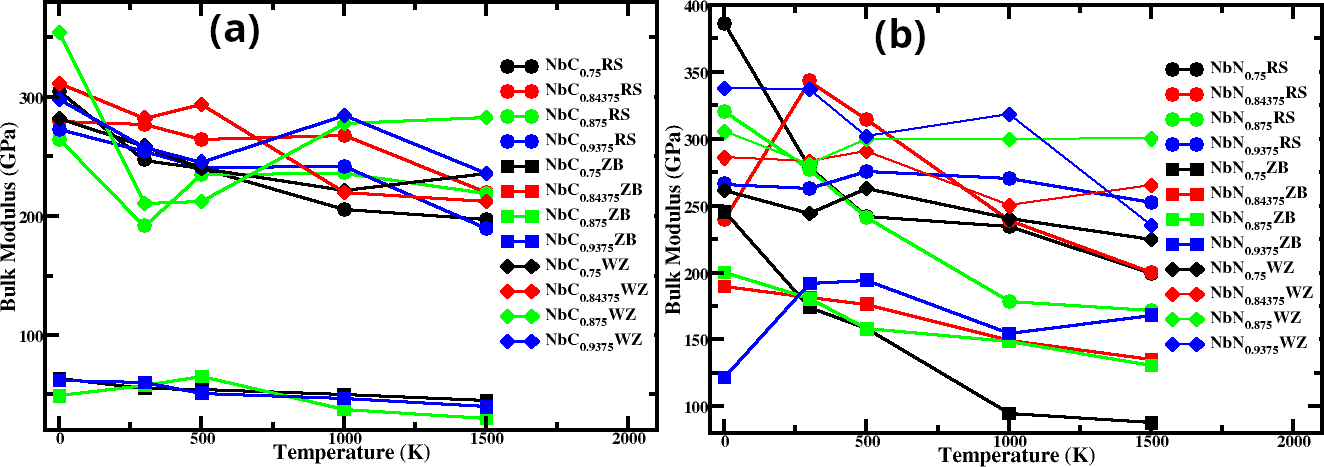}%
\caption{ Calculated bulk modulus (a) NbC and (b) NbN for RS, ZB and WZ for temperature range 0-1500 K. }
\label{NbC_NbN_BM}
\end{figure}

\begin{figure}[ht!]
\centering%
\includegraphics[width=1.1\linewidth]{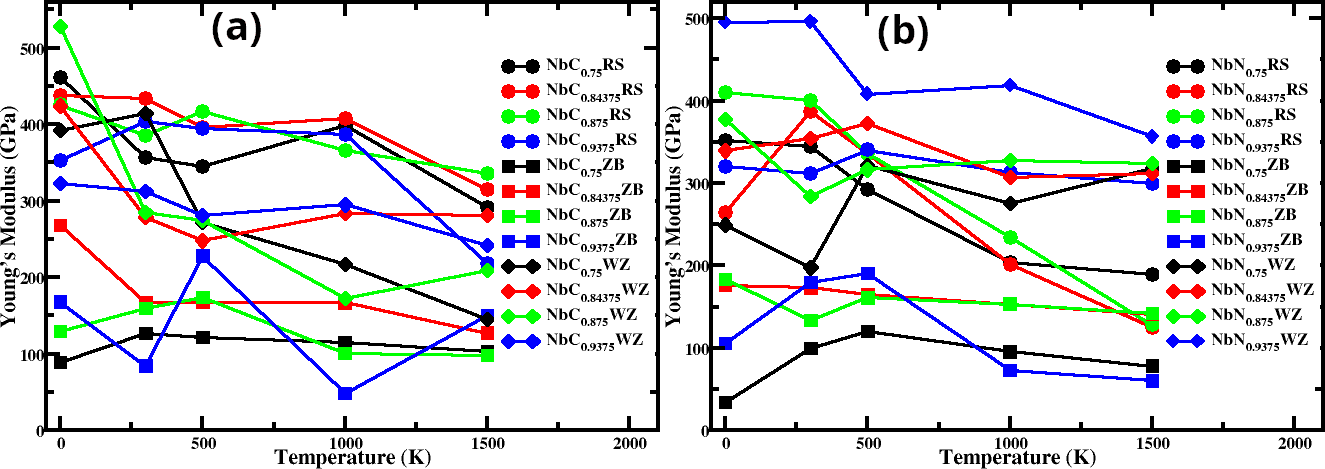}%
\caption{ Calculated Young's modulus (a) NbC and (b) NbN for RS, ZB and WZ for temperature range 0-1500 K. }
\label{NbC_NbN_YM}
\end{figure}

\begin{figure}[ht!]
\centering%
\includegraphics[width=1.1\linewidth]{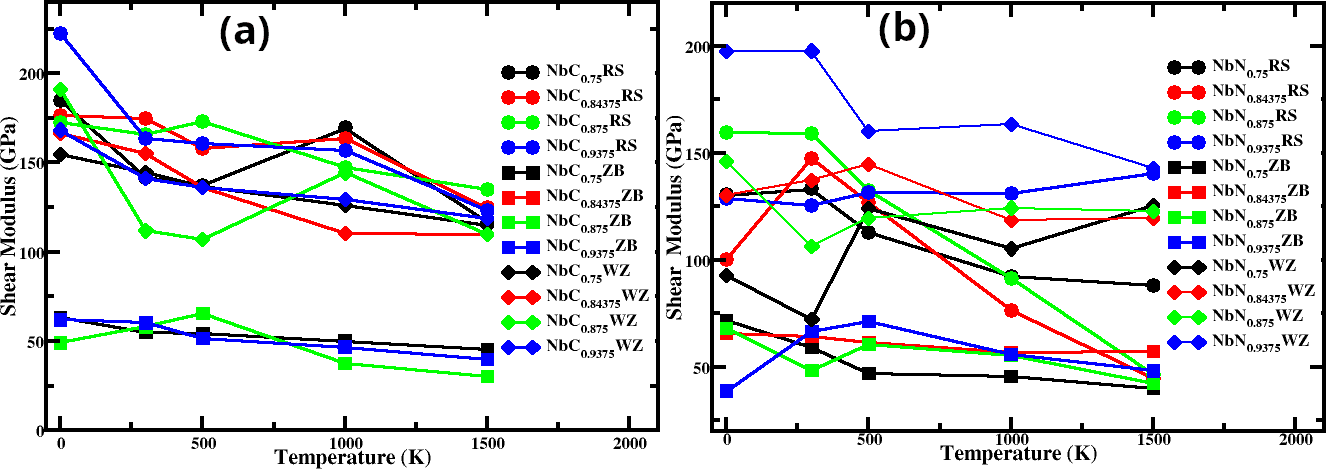}%
\caption{ Calculated Shear modulus (a) NbC and (b) NbN for RS, ZB and WZ for temperature range 0-1500 K. }
\label{NbC_NbN_SM}
\end{figure}

\begin{figure}[ht!]
\centering%
\includegraphics[width=1.1\linewidth]{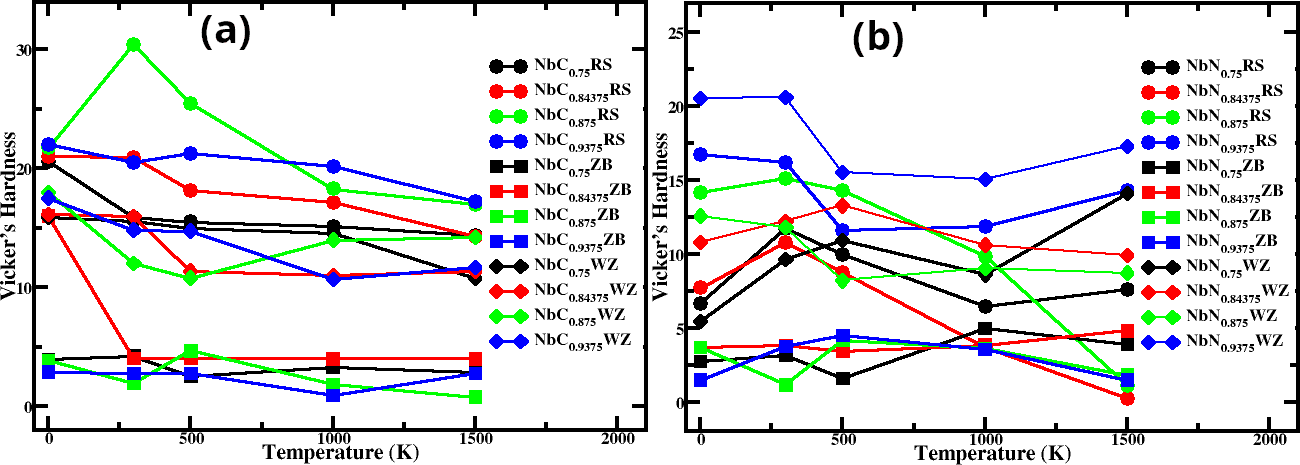}%
\caption{ Calculated Vicker's Hardness (a) NbC and (b) NbN for RS, ZB and WZ for temperature range 0-1500 K. }
\label{NbC_NbN_VH}
\end{figure}

The hardness of the NbC and NbN is calculated   using the equation below 
\begin{equation}\label{eq:Hv}
H_{V} = 2(K^{2}G^{0.585})-3
\end{equation}

where K= G/B being the Pugh’s ratio. The calculated mechanical properties are presented in Figure \ref{NbC_NbN_BM} – Figure \ref{NbC_NbN_VH}

It is observed that the bulk, Young’s and shear moduli of NbC and NbN decrease with increase in temperature for RS, ZB and WZ structures. For example, NbC$_{0.75}$ in RS decreases by 35 \%, 36 \%, 37 \% in Bulk, Young’s and Shear moduli, respectively, for  a temperature range 0-1500 K ( see Figure \ref{NbC_NbN_BM}(a)-\ref{NbC_NbN_SM}(a)). While NbN$_{0.75}$ in RS decreases by 48.32 \%, 46.23 \%, 32.25 \% in bulk, Young’s and shear moduli respectively for temperature range of 0 – 1500 K (see Figure \ref{NbC_NbN_BM}(b)-\ref{NbC_NbN_SM}(b)). Similarly, the values of bulk, Young’s and shear moduli decrease with increase in temperature for NbC$_{0.75}$ in ZB by 30.40 \%, 16.21 \%, 28.53 \%, respectively (see Figure \ref{NbC_NbN_BM}(a)-\ref{NbC_NbN_SM}(a) ) and by 64.17 \%, 41.65 \% and 43.98 \%, respectively, for NbN$_{0.75}$ in ZB (see Figure \ref{NbC_NbN_BM}(b)-\ref{NbC_NbN_SM}(b)). For NbC$_{0.75}$ in WZ, similar trend is observed as bulk, Young’s and shear moduli decrease by 16.37 \%, 63.04 \%, 25.79 \%, respectively with increase in temperature, as shown in Figure \ref{NbC_NbN_BM}(a)-\ref{NbC_NbN_SM}(a). While bulk, Young’s and shear moduli decrease by 13.94 \%, 38.65 \%, 42.46 \%, respectively, with increase in temperature from 0 – 1500 K for NbN$_{0.75}$ in WZ, as shown in Figure \ref{NbC_NbN_BM}(b)-\ref{NbC_NbN_SM}(b).
The progressive reduction in bulk, Young’s, and shear moduli is attributed to the weakening of Nb–C/N bonds with increasing temperature.  

\begin{figure}[ht!]
\centering%
\includegraphics[width=1.1\linewidth]{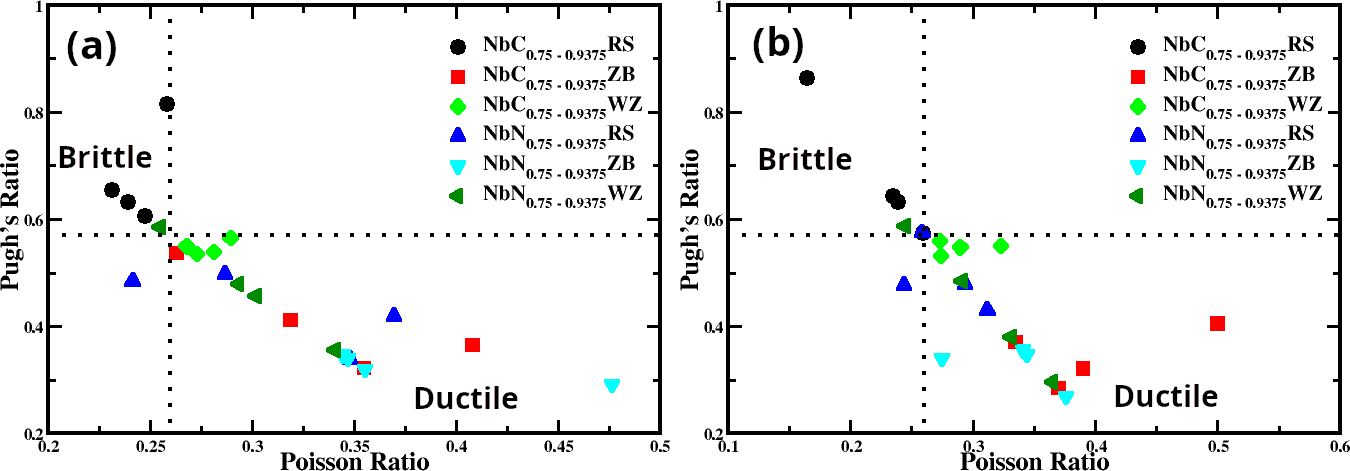}%
\caption{Toughness map for NbC and NbN for (a) 0 K (b) 300 K  with various defect concentrations.}
\label{0-300}
\end{figure}

\begin{figure}[ht!]
\centering%
\includegraphics[width=1.1\linewidth]{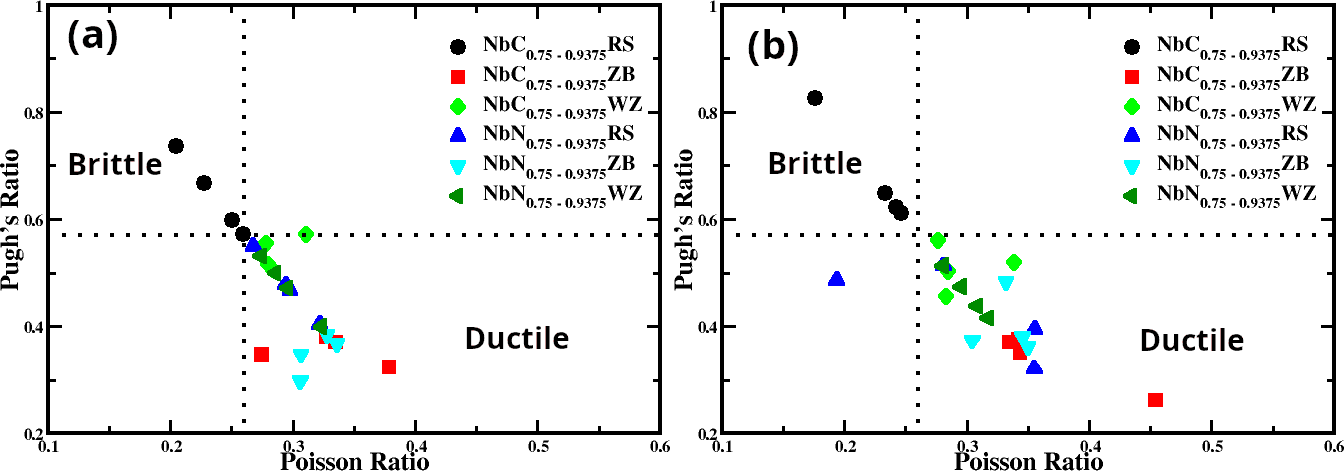}%
\caption{Toughness map for NbC and NbN for (a) 500 K (b) 1000 K  with various defect concentrations.}
\label{500-1000}
\end{figure}

\begin{figure}[ht!]
\centering%
\includegraphics[width=0.6\linewidth]{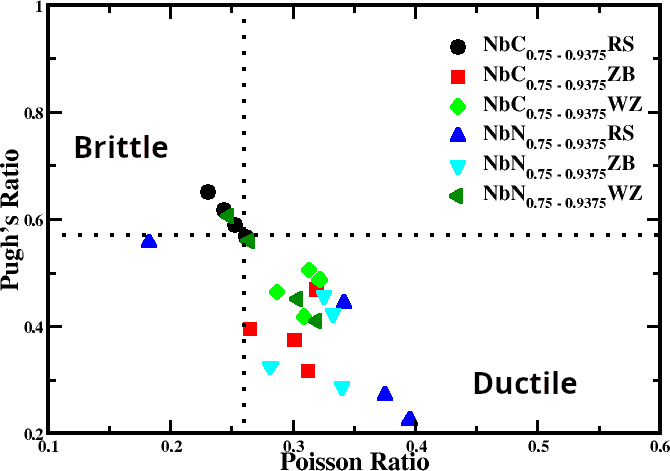}%
\caption{Toughness map for NbC and NbN for 1500 K  with various defect concentrations.}
\label{1500}
\end{figure}

Other defect concentrations were also considered and the same trend observed. The increase in defect concentration resulted in to decrease in the values of mechanical properties. For instance the bulk modulus, shear modulus and Young modulus of NbC/N$_{0.9375}$ decreased by 30.53/5.23 \%, 8.07/44.54 \% and 6.39/38.22 \%, respectively, as shown in Figure \ref{NbC_NbN_BM}-\ref{NbC_NbN_SM}. Similarly, for NbC/N$_{0.875}$ the trends in decrease in values of bulk modulus, shear modulus and Young’s modulus are as follows, 17.16/46.365 \%, 70.78/21.88 \%, and 68.76/21.08 \%, respectively, as shown in Figure \ref{NbC_NbN_BM}-\ref{NbC_NbN_SM}. While 21.34/16.535 \%, 27.09/52.55 \% and 29.35/55.275 \% decrease were observed in NbC/N$_{0.84375}$, as shown in as shown in Figure \ref{NbC_NbN_BM}-\ref{NbC_NbN_SM}. This behaviour is observed in the RS phase, moreover, a similar trend is also  noted in case of ZB and WZ, where  a reduction in the values of  the aforementioned parameters is observed.
Studies on the hardness characterization of niobium carbide and niobium nitride highlight a decrease in Vicker’s hardness with increasing temperature and defect concentration. For example, these materials possess high hardness values reflecting their superior mechanical stability and bonding strength. However, as temperature rises, the thermal agitation within the crystal lattice facilitates deformation, reducing the material's resistance to indentation. This is as a result of disturbance on the the bonding network, leading to a further decline in hardness. This trend was confirmed through density functional theory calculations performed in this study, showing that both NbC and NbN lose hardness as temperature and defects increase, which influences their suitability for hardness-related applications. For instance, in the RS crystal structure, the Vicker’s hardness of NbC/N$_{0.84375}$, NbC/N$_{0.875}$ and NbC/N$_{0.9375}$ decreased by 3.09/13.49 \%, 32.00/70.56 \% 21.84/9.20 \% and 29.69/14.6 \%, respectively. This provides critical information for understanding their behavior under operational conditions involving heat and lattice imperfections.\\ 
Furthermore, toughness analysis was performed based on Pugh,s criterion \cite{senkov2021generalization} where a value of  G/B $>$0.57  indicates brittleness and G/B$<$0.57 indicates ductility.  Additionally, Poisson ratio was used to assess the ductile and brittle nature of materials, with values greater than 0.26 corresponding to ductility and values below 0.26 indicating brittleness. By combining Pugh’s criterion with Poisson’s ratio, the toughness of NbC and NbN was mapped across different temperature ranges and defect concentrations. The toughness map for NbC and NbN in RS, ZB and WZ are presented for selected temperature within 0 - 1500 K temperature range.  At 0 K, NbC in the RS phase exhibits brittleness across all defect concentrations, whereas NbC in ZB and WZ, as well as NbN in RS, ZB, and WZ, are found to be ductile (Figure \ref{0-300}a). Increasing defect concentration promotes brittleness, as indicated by the rightward shift of the $B/G$ ratio (Figure \ref{0-300}b, Figure \ref{500-1000}(a–b), Figure \ref{1500}). Furthermore, the toughness maps ( see Figure \ref{0-300} - \ref{1500}) illustrate the influence of temperature on mechanical response, showing that ductility is progressively enhanced with increasing temperature and defect concentration.

\subsection{\textbf{Anionic Defect Barrier Energies} }
\begin{figure}[ht!]
\centering%
\includegraphics[width=0.6\linewidth]{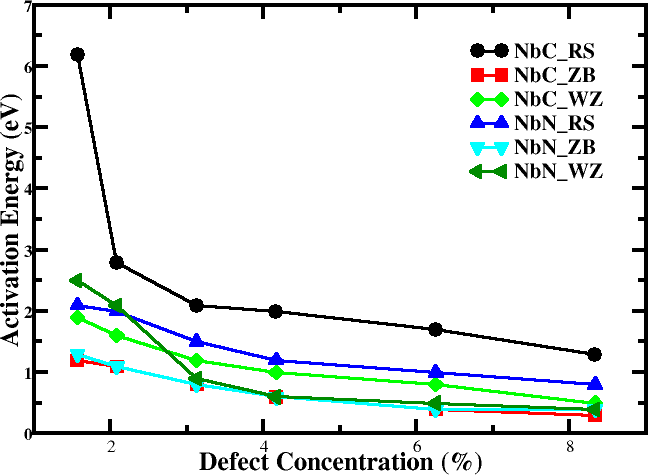}%
\caption{Calculated vacancy migration energy for Niobium Carbide together with Niobium Nitride in rocksalt, zincblende as well as wurzite crystal structures.}
\label{diffusion}
\end{figure}

Recent studies have shown that defects such as anionic vacancies can have a profound effect on the mechanical properties of NbC and NbN \cite{muchiri2022impact}. Thus, there is need for a comprehensive understanding of defect as it may lead to development of better control strategies \cite{angsten2014elemental}.  
In this work, we investigate anionic defect migration by calculating the migration energies of NbC and NbN in the RS, ZB, and WZ structures at various defect concentrations. The nudged elastic band (NEB) method \cite{sheppard2012generalized} was employed, using five discrete images between the initial and final states to map the defect migration path. Defect concentrations of 1.56 \%, 2.08 \%, 3.125 \%, 4.16 \%, 6.25\%, and 8.33 \% were considered. Migration energies were evaluated under fixed-cell geometries, where both the shape and size of the supercell were constrained during NEB minimization. Under these conditions, lower carbon concentrations correspond to smaller lattice constants and shorter bond distances, which facilitate faster atomic migration. To ensure consistency, defects were placed in the same lattice position for NbC and NbN in the RS, ZB, and WZ structures.
The results shows that migration barrier energies are strongly structure-dependent. For example, at a defect concentration of 1.56 \%, NbC exhibits migration energies of 6.2 eV, 1.3 eV, and 1.8 eV for the RS, ZB, and WZ phases, respectively ( as shown in Figure \ref{diffusion}). Furthermore, the vacancy migration energies of NbC and NbN in the same crystal structure are found to be very similar, reflecting the nearly identical atomic sizes of C and N. It is also observed that the barrier (activation) energies decrease with increasing defect concentration. The reduction arises from decreased repulsive interactions at larger interatomic spacings, which facilitates defect migration, as illustrated in Figure \ref{diffusion}.

\section{Conclusion}

This study systematically examined the combined influence of anionic vacancies and temperature on the mechanical properties of NbC and NbN using ab initio molecular dynamics and nudged elastic band calculations. The results demonstrate that elastic constants and moduli decrease progressively with rising temperature and vacancy concentration, reflecting the weakening of Nb–C/N bonds due to enhanced lattice vibrations and defect-induced distortions. Toughness mapping based on Pugh’s and Poisson’s criteria reveals that ductility generally improves at higher temperatures, while increasing defect concentrations promote brittleness. Vacancy migration energy calculations further confirm that defect mobility is strongly structure-dependent, with RS showing the highest energy barriers and WZ the lowest, and that NbC and NbN exhibit similar trends due to the comparable atomic sizes of carbon and nitrogen.  These findings emphasize the need to account for both temperature and defect effects when evaluating the mechanical performance of TMCNs. The insights gained here provide a fundamental framework for designing and optimizing NbC and NbN-based materials for demanding environments such as machining, aerospace, and high-temperature energy applications.

  \section{Acknowledgement}
 The authors gratefully acknowledge the Center for High Performance Computing (CHPC), South Africa, for providing the computational resources used in this research through MATS0868 and MATS862 accounts.



 \bibliographystyle{elsarticle-num} 
 \bibliography{cas-refs}





\end{document}